# Room-temperature tuning and probing of Fermi polarons in atomically thin semiconductors on a plasmonic metasurface

*Tingting Wu[1,#], Francesca Maria Marchetti[2,3,#], Antonio Tiene[2,3], Antonio I Fernández-Domínguez[2,3], Miao Qi[1], Zhe Wang[4], Lin Liu[1], Lei Wei[1], Francisco J. Garcia Vidal[2,3,*], Mengxiao Chen[5,*], Qi Jie Wang[1,*] and Yu Luo[6,*]*

AUTHOR ADDRESS

[1]School of Electrical and Electronic Engineering, Nanyang Technological University, Singapore 639798, Singapore

[2]Departamento de Física Teórica de la Materia Condensada, Universidad Autónoma de Madrid, Madrid 28049, Spain

[3]Condensed Matter Physics Center (IFIMAC), Universidad Autónoma de Madrid, Madrid 28049, Spain

[4]Key Laboratory of Bionic Engineering (Ministry of Education), Jilin University, Changchun 130022, China

[5]Department of Biomedical Engineering, Key Laboratory for Biomedical Engineering of Education Ministry, Zhejiang University, Hangzhou 310027, China

[6]National Key Laboratory of Microwave Photonics, Nanjing University of Aeronautics and Astronautics, Nanjing 211106, China

ABSTRACT: The Fermi polaron, arising from interactions between a mobile impurity and a degenerate Fermi sea, is a many-body quasiparticle that provides a sensitive probe of strongly correlated electronic phases in atomically thin semiconductors. In doped transition-metal dichalcogenides, the attractive and repulsive polaron branches are well established in monolayers. However, extending active control and quantitative, branch-resolved probing to stacked geometries has remained elusive because spectral quenching and weak optical contrast restrict access to Fermi polaron signatures. Here, we integrate electron-doped $WS_2$ flakes from monolayer to quadrilayer with a strain-tunable plasmonic metasurface, enabling high-contrast scattering readout at room temperature through coupling between Fermi polaron resonances and surface plasmons. This platform enables quantitative extraction of polaron branch spectral weights and coupling strengths across different layer numbers. We uncover a systematic thickness dependence of the spectral-weight distribution and demonstrate continuous and fully reversible spectral-weight transfer between attractive and repulsive branches in bilayers and quadrilayers, with near-complete transfer achieved in bilayers. By identifying layer number and strain as complementary control parameters for Fermi polarons, our results establish metasurface-enabled scattering spectroscopy as a practical route to resolve and manipulate many-body resonances in stacked van der Waals semiconductors, bridging idealized monolayer polaron physics and device-relevant architectures.

## INTRODUCTION

Transition metal dichalcogenides (TMDs) have emerged as a central platform for nanoscale science and technology[1,2], enabling major advances in nanoelectronics, optoelectronics, and quantum photonics. In the monolayer limit, their optical response is dominated by tightly bound excitons[3,4] arising from strong Coulomb interactions[4,5], while widely tunable charge doping establishes TMD monolayers as a unique testbed for light-matter interactions[6,7] and quantum many-body physics under device-relevant conditions[8]. A representative example is the Fermi polaron (FP)[9-11], a many-body quasiparticle formed when an optically generated exciton, acting as a mobile impurity, is dressed by many-body excitations of a surrounding degenerate Fermi sea. This gives rise to attractive (AP) and repulsive (RP) polaron branches that can be resolved in reflection or emission[9-13]. The FP is a paradigmatic problem in physics, appearing in systems ranging from ultracold atomic gases[14] to high-energy and condensed-matter settings[15]. In atomically thin TMDs, the FP framework provides a unified description of the optical response of doped monolayers, continuously bridging the AP and RP branches to neutral exciton and charged trion pictures, respectively, at low doping[16-20]. Branch-resolved access to these quasiparticles has enabled a broad range of interaction-driven phenomena, including boosted non-linearities[21], external-field control of dressed excitons[22,23], and optical signatures of correlated electronic states such as Wigner crystals[24,25], quantum Hall states[26], and correlated Mott states in Moiré superlattices[27].

In TMD monolayers, absorption and emission spectroscopies enable comprehensive characterization of the AP and RP branches[9,10,16,19,28-30]. However, extending active control and quantitative, branch-resolved probing to stacked geometries has remained elusive. This limitation is particularly consequential because multilayer TMDs represent the relevant operating regime for practical polaron-based technologies and are ubiquitous in optoelectronic and photonic devices[31-35]. Multilayer structures offer key practical advantages, including access to longer-lived interlayer excitations[36], additional tunability via stacking and dielectric screening[32,37,38], and enhanced light-matter interactions[39,40]. Yet the study of FPs in multilayers is complicated by their reduced emission intensity, broad spectral linewidths, and large screening, which exacerbate spectral quenching and obscure FP signatures in conventional absorption and photoluminescence (PL) measurements[41,42]. As a result, multilayer geometries generally lack a stable, branch-resolved spectroscopic handle that enables the deterministic control and quantitative tracking of the FP response at room temperature. Overcoming this bottleneck is therefore essential for translating polaron physics from monolayer testbeds to practical, reconfigurable optoelectronic platforms.

Here we introduce a nanophotonic strategy that provides a robust, branch-resolved optical readout of FPs at room temperature and enables their continuous and reversible control by uniaxial strain. We integrate polymer electron-doped $WS_2$ flakes from monolayer to quadrilayer onto a strain-tunable multi-singular plasmonic metasurface. Coupling between FP resonances and surface plasmons yields high-contrast scattering signatures even at room temperature, enabling the AP and RP branches to be resolved and quantitatively analyzed across repeated deformation cycles. Leveraging this capability, we demonstrate deterministic, continuous, and highly reversible strain control of the polaron response. Uniaxial strain induces a pronounced and repeatable redistribution of oscillator strength between the AP and RP branches, reaching near-complete spectral-weight transfer in bilayer $WS_2$. Furthermore, the accessible spectral-weight redistribution depends on the number of layers, establishing film thickness as a practical control parameter alongside strain. Together, these results establish metasurface-enabled scattering spectroscopy as a scalable route to resolve and program many-body resonances in doped van der Waals semiconductors, enabling reconfigurable polaronic optoelectronics under ambient conditions.

## RESULTS AND DISCUSSION

To achieve branch-resolved detection of FPs, we integrated $WS_2$ flakes ranging from monolayer (1L) to quadrilayer (4L) onto cold-etched, multi-singular gold (Au) metasurfaces supported by flexible conducting redox polymer (CRP) substrates[43] (Figure 1a). The metasurface hosts dense arrays of nanometer-scale gaps that support strongly confined gap-plasmon modes with large local-field enhancement (Figure 1b and Figure S2). This architecture yields a high-contrast scattering readout that enables room-temperature tracking of polaron-related spectral features, even in regimes where conventional PL becomes spectrally broadened and lacks the contrast required to resolve individual polaron branches (Figure 1e).

Electron doping in $WS_2$ is mediated by the CRP substrate, which acts as a tunable electron reservoir. Redox reactions in the CRP release free electrons[44] (Figure S1), while interfacial charge exchange with the Au metasurface raises the Au Fermi level, promoting electron injection into the adjacent $WS_2$ and establishing the carrier densities required for FP formation (Figure 1a, right)[9,10,45]. By varying the molecular weight of the CRP[43], the doping level can be systematically tuned across devices (Figure S1).

As illustrated schematically in Figure 1c, hybridization between the plasmonic modes and the FPs gives rise to three strongly coupled Fermi polaron polariton (FPP) branches, lower (L), middle (M), and upper (U), that exhibit characteristic anticrossing behavior as the plasmon resonance is tuned. These fingerprints are evident in the experimental spectra (Figure 1d; see Figure S4-S6 for systematic tracking of this spectral evolution with plasmon frequency, controlled via metasurface thickness). The M-FPP branch exhibits a unique spectroscopic hallmark of plasmon-mediated hybridization, remaining clearly visible even when the AP and RP resonances are closely spaced (Figure 1d, e). In contrast, such branch-resolved features are not accessible in standard PL measurements for multilayer $WS_2$ (Figure 1d, e): with increasing thickness, the PL peaks broaden and overlap while the emission intensity drops markedly (Figure 1d, right, inset), precluding reliable identification and tracking of the FP branches.

Notably, the dominant FP dip energies extracted from scattering in 2L and 4L $WS_2$ closely match those observed in 1L (Figure 1d), indicating a weak dependence on layer number. Rather than reflecting an absence of multilayer effects, this thickness invariance points to an effectively layer-selective optical response dominated by the metal-proximal layer. In this region, interfacial charge transfer and nanogap near-field coupling are strongest, while metal-induced dielectric screening establishes a comparable excitonic energy scale across different thicknesses[46]. As a consequence, contributions from layers farther from the interface are therefore suppressed, resulting in a selective readout of FP resonances that retains a single-layer-like energy structure within stacked geometries.

This metasurface-enabled spatial selectivity enables quantitative extraction of AP and RP spectral weights and coupling strengths across the 1L-to-4L series, despite the progressive quenching of PL with increasing thickness. The resulting scattering signatures are reproducible across multiple devices and remain stable over repeated measurement cycles, underscoring the robustness of the approach. Together, these results establish the metasurface-integrated scattering spectroscopy as an effective and practical route for room-temperature, branch-resolved access to FPs in multilayer $WS_2$, providing the basis for the strain-programmable polaron control demonstrated below.

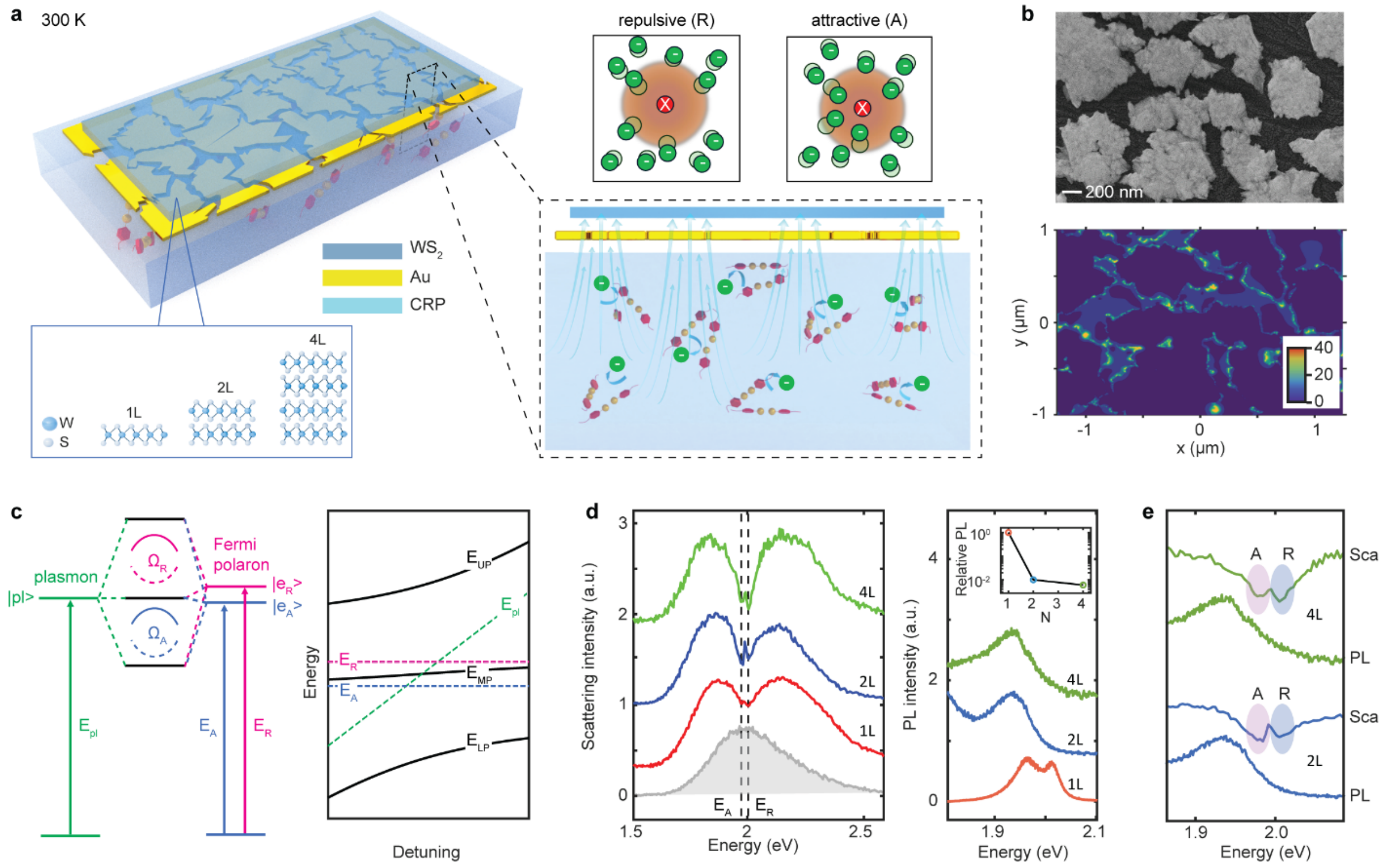

**Figure 1. Branch-resolved scattering readout of Fermi polarons in monolayer and few-layer $WS_2$ enabled by a plasmonic metasurface. (a) Left: Schematic of $WS_2$ flakes integrated with an Au metasurface on a flexible CRP substrate. The CRP acts as an electron reservoir; redox-mediated charge transfer through the Au enables electron injection into the metal-proximal $WS_2$, establishing the carrier density required for FP formation. (b) Scanning electron microscopy image of a 20-nm-thick Au metasurface (see Figure S2 for details) and simulated near-field intensity distribution of the gap plasmon modes. (c) Left: Conceptual coupling diagram for the plasmonic mode ($E_{\mathrm{pl}}$) interacting with the AP and RP resonances ($E_A$, $E_R$). Coupling to each branch is weighted by its spectral weight $Z_{A,R}$, yielding overall couplings $\Omega_{A,R} = \Omega\sqrt{Z_{A,R}}$ and producing three hybrid branches (L, M, U). Subscripts $A$ and $R$ denote the AP and RP branches, respectively. Right: Schematic evolution of the coupled modes with detuning $E_{pl} - E_R$. (d) Left: Dark-**

**field scattering spectra (arbitrary units, a.u.) of monolayer and multilayer $WS_2$ integrated with the plasmonic metasurface. The gray-filled curve denotes the bare plasmonic resonance measured in the absence of $WS_2$. Vertical dashed lines indicate the extracted FP energies $E_A$ and $E_R$. Right: PL spectra of electron-doped $WS_2$ on CRP for different layer numbers. Inset: normalized PL intensity versus layer number *N*. (e) Representative magnified comparison of scattering (Sca) and PL spectra for 2L and 4L $WS_2$.**

We first use 1L $WS_2$ as a benchmark to validate a quantitative framework that links the metasurface-enabled scattering readout to intrinsic FP parameters. In 1L TMDs, a well-established microscopic theory[19] rigorously connects PL emission and absorption spectra, enabling the extraction of fundamental FP parameters, including energies ($E_{A,R}$), spectral weights ($Z_{A,R}$), and linewidths ($\Gamma_{A,R}$), from conventional PL measurements. These PL-derived quantities are obtained independently of the metasurface scattering analysis and therefore serve as an external benchmark for testing the optical readout model. This independent determination of the FP response provides a stringent test of our scattering model: by fixing these key FP parameters a priori, we directly test whether an independently established FP response can reproduce the measured scattering lineshapes with minimal fitting freedom. Establishing this quantitative link is essential before applying the readout to stacked samples, where PL becomes spectrally broader and weaker, and loses its reliability as a branch-resolved probe (Figure 1d,e).

Monolayer $WS_2$ exhibits exceptionally strong Coulomb interactions, with reported exciton binding energies in the range of ~200–700 meV, depending on the dielectric environment and fabrication conditions[4,5]. These values are significantly larger than the room-temperature thermal energy ($k_B T \approx 26$ meV). As a result, excitonic resonances in $WS_2$ remain robust under ambient conditions, providing an ideal setting for the many-body interaction and dressing of excitons by an electron gas.

For the 1L benchmark, we use a CRP substrate (weight-average molecular weight 92,343 g/mol; Figure S1) to provide electron doping and resolve both FP branches in PL. The PL spectra exhibit two well-separated peaks over a wide temperature range (Figure 2a), assigned to the AP and RP branches. The temperature-dependent energy splitting agrees with theoretical predictions from the microscopic model of doped 1L TMDs[19] (Figure 2b). The redistribution of PL intensity from AP to RP with temperature follows Boltzmann-type scaling between emission $P(\omega)$ and absorption $A(\omega)$ (Figure S3), governed by the detailed-balance relation[47]:

$$P(\omega) = e^{-\frac{\omega}{k_B T}} A(\omega) \tag{1a}$$

$$A(\omega) = -\frac{1}{\pi} Im[G_X^0(\omega)], \tag{1b}$$

where $G_X^0(\omega)$ is the exciton Green's function

$$G_X^0(\omega) = [\omega - \Sigma(\omega)]^{-1}, \tag{2}$$

and $\Sigma(\omega)$ is the exciton self-energy arising from exciton-electron interactions (see Ref.[19] for details). By comparing the measured AP–RP splitting with this finite-temperature microscopic theory, we first determine the electron doping level ($E_F \sim 5\ meV$, Figure 2c). Once $E_F$ is fixed, the corresponding bare FP parameters ($E_{A,R}$, $Z_{A,R}$, $\Gamma_{A,R}$) follow directly from the theory and are then used as fixed inputs for the scattering analysis. Note that, following the sum rule of the absorption (1b), $\int_{-\infty}^{+\infty} d\omega\ A(\omega) = 1$, one can show that the spectral weights satisfy $Z_A + Z_R \sim 1$ independent of doping and temperature. This is because the trion-hole continuum has either a negligible weight or, at low doping or high temperature, is subsumed into the attractive branch[19].

Integrating 1L $WS_2$ with the Au metasurface, the dark-field scattering spectra (Figure 2d) exhibit pronounced spectral features arising from coupling between the plasmonic resonance and the FP response. To quantitatively reproduce these lineshapes, we model the measured scattering intensity $S(\omega)$ as the sum of three physically distinct contributions:

$$S(\omega) = \left|G_{pl}(\omega)\right|^2 + x\left|G_{pl}^0(\omega)\right|^2 - yA(\omega), \tag{3}$$

The dominant first term, $\left|G_{pl}(\omega)\right|^2$, captures scattering from the hybrid FPP response through the coupled plasmon Green's function

$$G_{pl}(\omega) = [\omega - E_{pl} + i\eta_{pl} - (\Omega/2)^2 G_X^0(\omega)]^{-1}. \tag{4}$$

Here, $E_{pl}$ and $\eta_{pl}$ are the plasmon energy and damping rate, respectively, and $\Omega$ is the Rabi coupling strength. The second term accounts for scattering from bare uncoupled plasmons, and the third represents absorption by bare non-hybridized FPs; $x$ and $y$ parameterize their relative weights.

In Eq. (3), the exciton Green's function (2) can be simplified as

$$G_X^0(\omega) \sim \frac{Z_R}{\omega - E_R + i\Gamma_R} + \frac{Z_A}{\omega - E_A + i\Gamma_A} \tag{5}$$

where FP energies $E_{A,R}$, spectral weights $Z_{A,R}$, and half-linewidths $\Gamma_{A,R}$ are taken directly from the PL-based microscopic analysis (Figure 2c). The plasmon parameters ($E_{pl}$, $\eta_{pl}$) are extracted from the bare Au metasurface. Thus, the scattering spectra can be quantitatively reproduced by fitting a minimal number of parameters, i.e., the coupling strength $\Omega$ and the residual weighting factors $x$ and $y$. The fitted value of $\Omega$ is compatible with the value obtained independently by employing a coupled oscillator model (Figure S6). Crucially, the intrinsic FP parameters are not treated as adjustable variables but are fixed from independent PL measurements.

The model reproduces the experimental spectra with excellent agreement (Figure 2d, representative fit parameters $x = 0.75$, $y = 0.04$, $E_A = 1.972\ eV$, $E_R = 2.01\ eV$, $\Gamma_A = 50\ meV$ and $\Gamma_R = 35\ meV$), with the observed anticrossing corresponding to a coupling strength of $\Omega\ =$ $284 \pm 11$ meV. Robustness is confirmed by tuning the plasmon resonance via Au thickness (Figure 2d) and by repeating the analysis using a second CRP substrate with lower molecular weight (31,386 g/mol), which increases the doping level ($E_F$~9 meV; Figure S1 and S5). In all cases, the scattering spectra are quantitatively captured using FP parameters determined independently benchmarked from the 1L PL spectroscopy, rather than by freely re-fitting microscopic FP parameters for each sample.

This internally consistent validation establishes metasurface-enabled scattering as a reliable, high-contrast method for extracting and tracking FP branch energies and spectral weights at room temperature. In the following sections, we use the same validated framework to analyze stacked $WS_2$ samples, where conventional PL is insufficient for branch-resolved tracking, and to quantify strain-programmable polaron control.

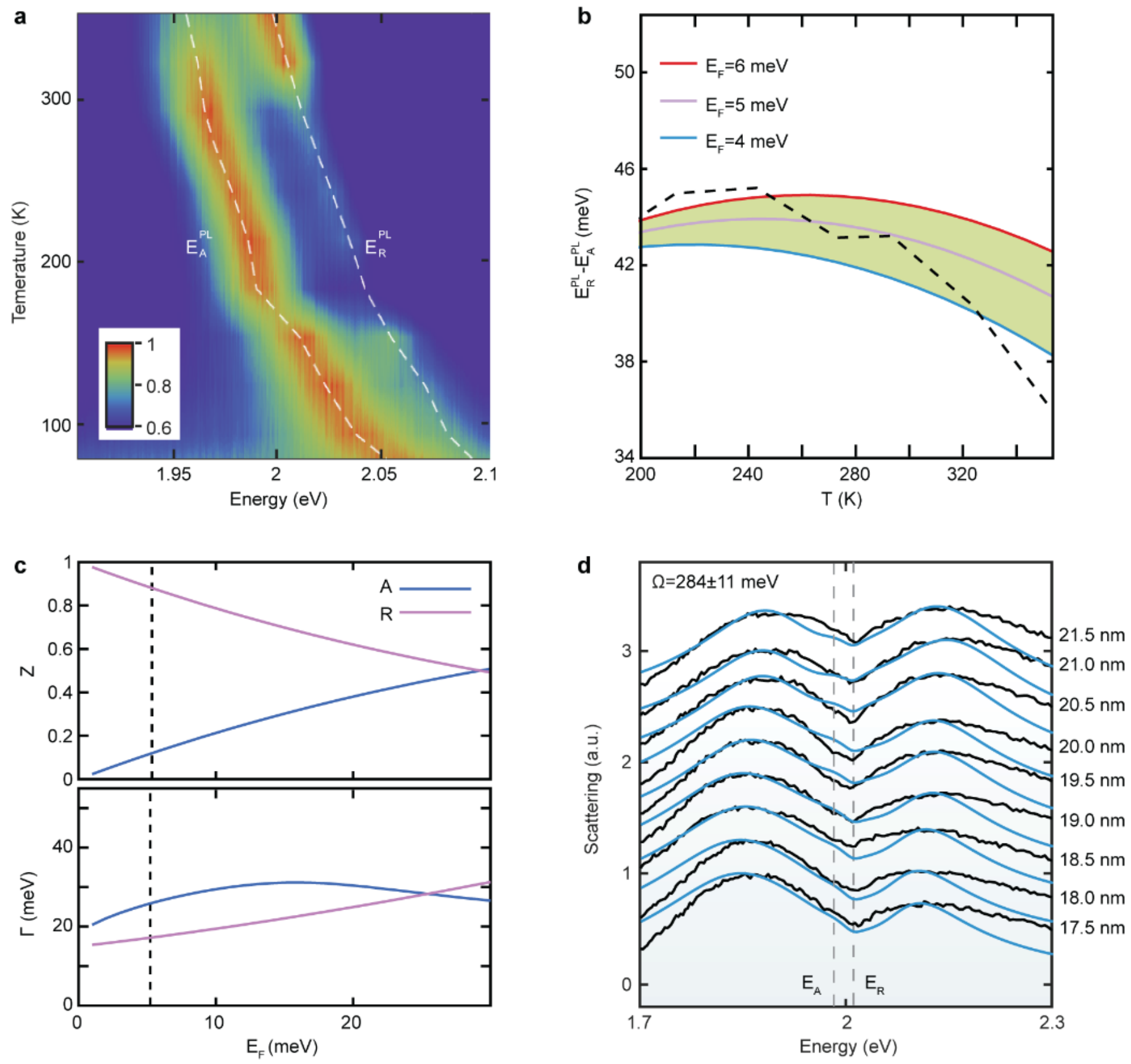

**Figure 2. Benchmarking monolayer $WS_2$ Fermi polarons and validating the metasurface-enabled scattering readout. (a) Temperature-dependent PL emission maps of electron-doped monolayer $WS_2$ on a CRP substrate, showing resolved AP and RP peaks (white dashed guides). (b) Temperature dependence of the energy splitting between AP and RP peaks, $E_R^{PL} - E_A^{PL}$, extracted from PL spectra (dashed line), compared with microscopic theory calculations for selected Fermi energies $E_F$ (solid lines). (c) Calculated FP spectral weights $Z_{A,R}$ and half-linewidths $\Gamma_{A,R}$ as functions of $E_F$; the vertical dashed line marks the $E_F$ determined by matching the measured PL splitting in (b) (Methods). (d) Dark-field scattering spectra of monolayer $WS_2$ integrated with an Au metasurface on a CRP substrate for different Au film thicknesses (17.5-21.5 nm, bottom to top), which tune the plasmon resonance from 1.96 eV to 2.028 eV and hence the plasmon-polaron detuning**

**(Figure S4). Black curves show the measured spectra, while blue curves show model calculations using FP parameters independently determined from the PL benchmark in (a)-(c) and plasmonic parameters extracted from the bare metasurface (Methods and Figure S4). Vertical dashed lines indicate the FP energies $E_{A,R}$ in scattering (note that $E_{A,R}^{\mathrm{PL}}$ in (a) are temperature-dependent and therefore not identical to the room-temperature dip positions in (d)).**

With the scattering model validated on 1L $WS_2$, we now apply the same framework to multilayer samples (2L and 4L). In contrast to 1L, conventional PL from bare 2L and 4L $WS_2$ does not provide a reliable branch-resolved FP readout because the emission becomes broadened and spectrally weaker (Figure 1d, right). When the flakes are integrated onto the Au metasurface supported by a CRP substrate, the dark-field scattering spectra display three hybrid branches, L-, M-, and U-FPPs, together with two pronounced dips that mark the AP and RP resonance energies $E_{A,R}$ (Figure 3a,b).

We analyze the multilayer spectra using the same framework established for 1L $WS_2$, extracting the FP properties directly from the dark-field scattering measurements. The FP energies $E_{A,R}$ and plasmon parameters $(E_{pl}, \eta_{pl})$ are fixed, leaving four parameters to be determined: the attractive polaron spectral weight $Z_A$ (or equivalently $Z_R$), the Rabi splitting $\Omega$, and the residual weights of the uncoupled modes, $x$ and $y$. To ensure robustness, the extracted value of $\Omega$ is cross-checked using an independent coupled-oscillator model. As detailed in the Supporting Information, the total coupling strength can be written as $\Omega = \sqrt{\Omega_A^2 + \Omega_R^2}$, with $\Omega_{A,R} = \Omega\sqrt{Z_{A,R}}$, linking the microscopic FP parameters to the oscillator model. Importantly, the FP linewidths $\Gamma_{A,R}$ are not treated as free fitting parameters; instead, they are varied within experimentally consistent ranges, and the resulting uncertainty is propagated to assign error bars to the extracted $Z_A$.

The extracted bare FP energies remain nearly unchanged from 1L to 2L and 4L (Figure 3a,b). This thickness invariance suggests that the detected branch-resolved FP signatures retain an effectively single-layer-like character in stacked geometries. As mentioned above, a plausible explanation is that the response is dominated by the metal-proximal layer, where charge transfer and nanogap near-field coupling are strongest and where metal-induced dielectric screening sets a similar excitonic energy scale across thickness[46,48]. Accordingly, layer number is treated here primarily as an experimental control parameter that reshapes the metasurface readout and the accessible redistribution of branch spectral weights, rather than invoking a fully layer-resolved multilayer FP model.

Consistent with this view, the FP-related dips in 2L and 4L do not exhibit strong additional broadening relative to 1L. This behavior is inconsistent with a simple vertical-averaging picture in which unresolved contributions from multiple inequivalent layers would yield inhomogeneous broadening. Instead, it indicates an effectively layer-selective nanogap readout that suppresses contributions from layers farther from the interface, where both optical coupling and effective doping are reduced. Stacked geometries can also partially shield the optically active, metal-proximal region from ambient adsorbates and local charge fluctuations, further reducing inhomogeneous broadening.

With increasing layer number, the L- and U-FPP branch energies shift in opposite directions, while the M-FPP branch remains between the closely spaced FP resonances (Figure 3a,b, representative fit parameters $x = 0.9$, $y = 0.3$, $E_A = 1.967\ eV$, $E_R = 2.003\ eV$, $\Gamma_A = 50\ meV$ and $\Gamma_R = 40\ meV$). From these fits we extract the FP spectral weights $Z_{A,R}$ (Figure 3d). The effective coupling strength $\Omega$ is compatible with the value obtained by the experimentally observed anticrossing splitting (Figure 3c and Figures S6, S7). $\Omega$ increases from $284 \pm 11$ meV (1L) to $326 \pm 10$ meV (2L) and $350 \pm 10$ meV (4L) (Figure 3c). The increase in $\Omega$ with layer number deviates from ideal $\sqrt{N}$ scaling, consistent with non-uniform vertical coupling and doping profiles (Figure S9), such that only layers within the near-field decay length and with appreciable carrier density contribute effectively. In bilayer $WS_2$, the second layer can acquire carriers via interlayer transfer while remaining within the near-field decay length, producing a larger overall coupling, whereas in quadrilayers the near-field amplitude and carrier density decay with distance from the interface, so the net increase in $\Omega$ is modest but reproducible.

A pronounced thickness dependence is also observed in the extracted branch spectral weights $Z_{A,R}$ (Figure 3d). The response evolves from an RP-dominated regime in 1L ($Z_A = 0.13$, $Z_R = 0.87$) to an approximately balanced distribution in 2L ($Z_A = 0.485 \pm 0.06$, $Z_R = 0.515 \mp 0.06$), followed by a partial reversion in 4L ($Z_A = 0.45 \pm 0.07$, $Z_R = 0.55 \mp 0.07$). The quoted uncertainties for 2L and 4L primarily reflect the fitting sensitivity to the refined broadening ratio $\Gamma_A/\Gamma_R$. This systematic evolution indicates that the relative branch contribution depends sensitively on layer number.

We attribute this behavior to thickness-dependent changes in the effective electron density and its vertical distribution in stacked geometries, which are expected to modify the branch spectral weights sampled by the near-field readout. In 1L under ambient conditions, adsorption of $H_2O$ and $O_2$ can reduce the steady-state electron density[49,50], suppressing the AP weight and enhancing inhomogeneous broadening. Although encapsulation with hexagonal boron nitride (hBN) can stabilize doping in monolayer TMDs, it is not compatible with the mechanically reconfigurable, metasurface-integrated architecture employed here, as additional interfaces impede repeatable strain transfer and reduce scattering contrast over multiple deformation cycles. Few-layer $WS_2$ therefore constitutes a practical and functionally distinct geometry, providing partial environmental shielding while preserving strong coupling to the plasmonic near field and mechanical compatibility with cyclic strain control. In this regime, thicker stacks can sustain higher effective carrier densities within the optically active region, while distributing carriers over additional layers from 2L to 4L reduces the carrier density per interfacial layer. This redistribution progressively suppresses the evolution of the AP spectral weight $Z_A$ with increasing thickness, without appreciably shifting the FP energies, which remain nearly pinned at $E_{A,R}$.

We note that the thickness dependence of $\Omega$ (Figure 3c) is less pronounced than in undoped multilayer $WS_2$[39], consistent with reports that the AP acquires an out-of-plane dipole component[37,51] that couples less efficiently to predominantly in-plane plasmonic fields. In this situation, an increased AP spectral weight can partially offset the coupling gain expected from adding layers.

Overall, our multilayer data can be summarized by two experimentally robust observations: (i) the overall coupling strength $\Omega$ increases with layer number, $N$, but deviates from ideal $\sqrt{N}$ scaling, consistent with non-uniform vertical coupling and doping profiles; and (ii) the extracted branch spectral weights $Z_{A,R}$ exhibit a pronounced thickness dependence. Together, these trends show that thickness provides a practical degree of freedom that reshapes the accessible branch spectral weights and the strength of nanophotonic hybridization in stacked $WS_2$, which we combine with strain tuning in the following section.

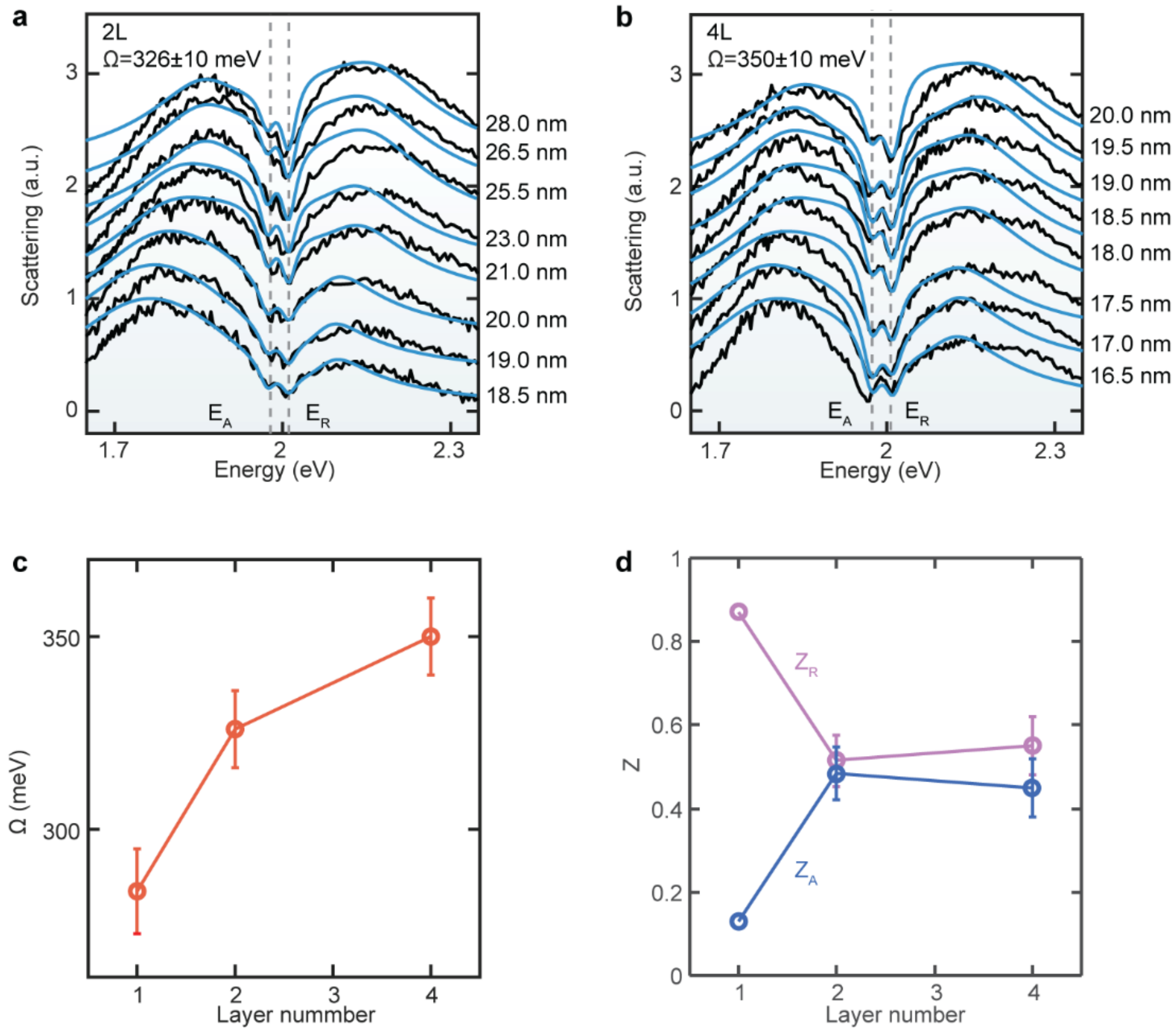

**Figure 3. Fermi polarons in multilayer $WS_2$. Dark-field scattering spectra of 2L (a) and 4L (b) $WS_2$ on Au metasurfaces supported by a CRP substrate, measured for different Au film thicknesses (bottom to top: 18.5-28 nm for 2L and 16.5-20 nm for 4L), which tune the plasmon resonance and hence the detuning. The spectra display two broad peaks (L-FPP and U-FPP) and a narrower central peak (M-FPP). Vertical dashed lines indicate the FP energies $E_{A,R}$. Black and blue curves correspond to measured and calculated spectra, respectively. (c) Effective coupling strength Ω as a function of $WS_2$ layer number, extracted from the experimentally observed anticrossing splitting of the hybrid FPP branches (Figure S7). (d) Effective optical branch weights $Z_{A,R}$ as a function of $WS_2$ layer number, obtained from the branch-dependent coupling strengths in the three-branch FPP model.**

**Error bars for Ω indicate variation across different Au thicknesses, while those for $Z_{A,R}$ reflect fitting uncertainties associated with variations in the polaron linewidths.**

Active control of excitonic quasiparticles in two-dimensional semiconductors is essential for exciton-based optoelectronic and polaritonic platforms, where tunable light-matter interactions underpin energy transduction and photonic functionality[52,53]. Here, using the metasurface-enabled, branch-resolved scattering readout established above, we demonstrate that uniaxial mechanical deformation provides a deterministic and reversible knob to redistribute oscillator strength between the AP and RP in $WS_2$ flakes under ambient conditions (Figure 4).

Uniaxial strain continuously modulates near-field optical confinement and plasmon-assisted hot-carrier generation by reshaping the geometry and hotspot density of the metasurface nanogaps. To efficiently drive this process, we employ a supercontinuum white-light laser[54] to resonantly excite the localized plasmon modes, replacing the broadband lamp illumination used previously. This targeted excitation significantly enhances hot-carrier generation in the Au metasurface, establishing an optically controlled pathway to tune the local electron density in the adjacent $WS_2$[55,56]. Consequently, strain-controlled variations in hotspot strength and hot-electron injection directly redistribute oscillator strength between the AP and RP branches, leveraging the high sensitivity of the FP spectral weights to carrier density.

Experimentally, varying strain produces pronounced and systematic changes in the scattering spectra of both monolayer and multilayer $WS_2$ integrated with the metasurface (Figure 4b for 2L and 4L; Figure S8 for 1L). Tensile strain produced by upward bending widens the nanogaps, reduces field confinement, and suppresses plasmon-assisted hot-electron injection, lowering the effective doping (Figure 4a). Conversely, compressive strain induced by downward bending narrows the gaps, strengthens field confinement, and enhances plasmon-assisted hot-electron generation and injection[55,57,58], increasing the effective doping. As a result, tuning the strain from -0.2% to +0.2% drives a near-complete redistribution of oscillator strength between the two FP branches in few-layer samples: for 2L, the AP (RP) weight changes from 0.81 (0.19) to 0.17 (0.83); for 4L, it changes from 0.76 (0.14) to 0.31 (0.79). Over the same strain range, the conversion is less complete in monolayer $WS_2$ (Figure S8, AP (RP) weight changes from 0.2 (0.8) to 0.02 (0.98)), consistent with reduced and less stable effective doping in exposed monolayers under ambient adsorption[49,50].

In both 2L and 4L $WS_2$, the dominant effect of strain is a pronounced and reversible redistribution of spectral weight between the AP and RP branches, without any appreciable rigid shift of their energies. This behavior is consistent with carrier-density modulation, as finite-temperature FP theory predicts that the AP/RP spectral weights are highly sensitive to carrier density. If strain were to induce substantial band shifts, corresponding changes in the resonance energies would be expected. Instead, the peak positions remain nearly unchanged throughout the entire deformation range. Consistently, control PL measurements on monolayer $WS_2$ decoupled from the Au metasurface using a weak-coupling spacer structure (20 nm $Al_2O_3$ inserted between $WS_2$ and the Au metasurface; Figure S10), where plasmon–polaron coupling and hot-electron injection are suppressed, exhibit only modest peak-energy shifts and minor linewidth variations under comparable strain, without any evidence of branch-like spectral-weight transfer. These observations indicate that strain-induced band-edge shifts and the associated renormalization of FP energies and linewidths play only a secondary role under the present conditions, whereas the dominant mechanism originates from strain-controlled plasmonic confinement and hot-electron injection.

In multilayers, vertical charge redistribution together with protection of the bottom layer from surface adsorbates amplifies the sensitivity of the effective FP spectral weights to changes in hot-electron doping, enabling a substantially broader tuning range than in the monolayer limit. This enhanced tunability arises from the combined effects of plasmonic field distribution and layer-dependent FP characteristics, which jointly govern the layer-selective carrier injection and the resulting branch spectral weights. Specifically, at small nanogap sizes, the strongly confined plasmonic field generates a high density of hot electrons that preferentially dope the $WS_2$ layers closest to the Au interface, favoring the formation of AP and increasing $Z_A$. As the gap size increases under tensile strain, plasmonic field confinement weakens, and the relative contribution of $WS_2$ layers farther from the Au surface becomes more significant. Because the AP spectral weight decreases systematically with increasing layer index, the vertically averaged response shifts toward the repulsive branch, resulting in a reduced effective $Z_A$. This layer-resolved physical picture provides a consistent mechanism for the observed broadening of the tuning range in few-layer systems.

Despite the pronounced strain-induced redistribution of FP spectral weights, the overall Rabi coupling strength Ω remains nearly invariant with strain (Figure 4c). Strain therefore acts primarily to reallocate oscillator strength between the FP branches (Figure 4d) rather than to modify the net coupling strength. This behavior is consistent with a compensation mechanism: compressive strain increases plasmonic field confinement but also increases the spectral weight of the AP, which acquires an out-of-plane dipole-moment component[37,51] and couples less efficiently to predominantly in-plane plasmonic fields. Tensile strain weakens confinement but increases the spectral weight of the RP, which couples more efficiently to in-plane fields. These opposing trends stabilize Ω across the entire strain range.

The reversible and near-complete conversion between repulsive- and attractive-branch dominance demonstrated here establishes uniaxial strain as a precise control knob for engineering many-body quasiparticles in TMDs at room temperature. By enabling deterministic and repeatable redistribution of oscillator strength between FP branches over deformation cycles while maintaining a nearly constant coupling strength, this approach provides a practical route toward adaptive and reconfigurable polaritonic functionalities in mechanically tunable nanophotonic platforms.

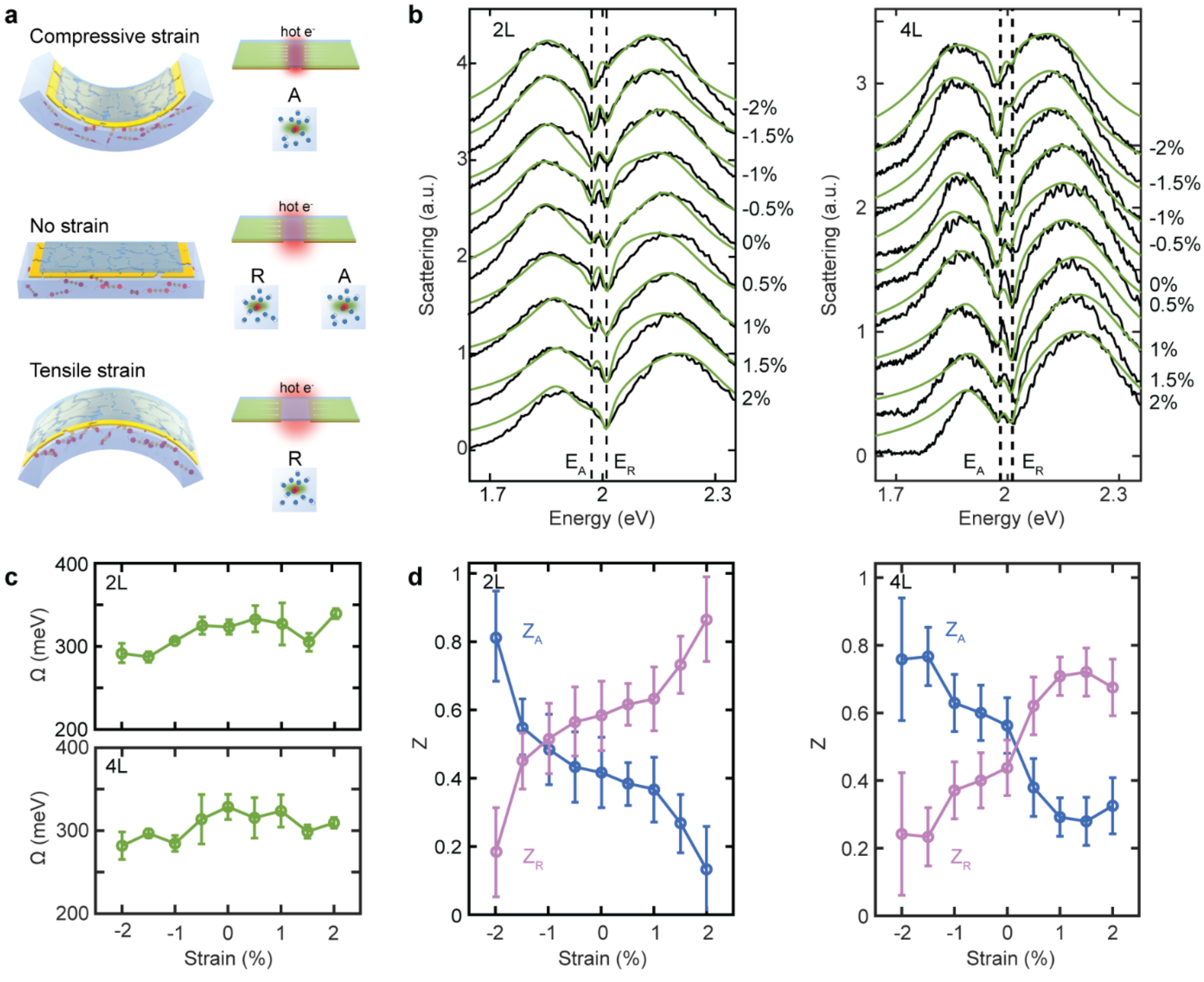

**Figure 4. Strain-programmable redistribution of Fermi polarons. (a) Schematic of uniaxial strain control in the $WS_2$-plasmonic platform. Upward (tensile) bending widens the**

**nanogaps in the metasurface, reduces near-field confinement and plasmon-assisted hot-electron injection, and shifts oscillator strength toward the RP branch. Downward (compressive) bending narrows the gaps, enhances near-field confinement and hot-electron injection into $WS_2$, and shifts oscillator strength toward the AP branch. (b) Normalized dark-field scattering spectra for 2L (left; Au thickness 22 nm) and 4L (right; Au thickness 19 nm) $WS_2$ plasmonic systems under different applied strains (2% to -2%, bottom to top). Measured spectra are shown in black, and calculated spectra in green. Vertical dashed lines indicate the FP energies $E_{A,R}$. Rabi splitting $\Omega$ (c) and effective FP spectral weights $Z_{A,R}$ (d) as functions of strain for 2L (top/left) and 4L (bottom/right) $WS_2$. Error bars reflect fitting uncertainties due to variations in the polaron linewidths.**

## CONCLUSIONS

We have established a practical platform for resolving and manipulating Fermi polarons at room temperature. By integrating polymer-doped $WS_2$ flakes with a strain-tunable multi-singular plasmonic metasurface, we realize a high-contrast scattering readout that separates attractive and repulsive branches under ambient conditions, including regimes where the modes become spectrally unresolvable in photoluminescence. Using monolayer $WS_2$ as a benchmark, we validate a parameter-free framework that links scattering lineshapes to intrinsic polaron parameters, enabling systematic and quantitative measurements across devices.

Extending this readout to few-layer stacks, we uncover a dependence of the accessible branch spectral weights on layer number. Leveraging this capability, we demonstrate deterministic and fully reversible strain programming of oscillator-strength distribution between branches over repeated deformation cycles, achieving near-complete repulsive-to-attractive conversion in bilayers. The observed programmability is consistent with strain-controlled modulation of plasmonic confinement and optically assisted hot-carrier injection, which together enable large and repeatable reconfiguration of branch spectral weights.

More broadly, our results show that metasurface-enabled nanophotonic scattering provides a scalable route to resolve and control many-body resonances in doped van der Waals materials under ambient conditions. The combination of a high-contrast optical readout with mechanically tunable hot-carrier injection should be transferable to other material systems and nanophotonic architectures, opening opportunities for reconfigurable many-body optoelectronics and polaritonic devices. Looking ahead, extending this approach to a wider range of quasiparticles could enable adaptive photonic circuits that actively harness and reconfigure correlated optical responses on demand.

## EXPERIMENTAL METHODS

**Sample Preparation.** The gold plasmonic metasurface was developed using a cold-etching technique. $WS_2$ flakes were mechanically exfoliated from bulk crystals and subsequently transferred onto the gold metasurface through a dry transfer technique. The layer number of the $WS_2$ flakes was determined by PL spectra and optical contrast. Electron doping of the $WS_2$ flakes was achieved and regulated using a CRP.

**Optical Measurements.** All optical measurements were conducted at room temperature in a reflective configuration. PL measurements were performed using a 532 nm diode laser focused through a 100× microscope objective with a numerical aperture (NA) of 0.75. The excitation power was approximately 100 μW with full width at half maximum (FWHM) beam diameter of ~1 μm. Dark-field scattering measurements were carried out using two light sources. Hyperspectral imaging was first performed with a broadband halogen lamp. Subsequently, a supercontinuum white-light laser was employed to excite plasmonic hot electrons. In both cases, the incident light was focused using a 50× objective (NA = 0.55) with an excitation power of 20 μW, yielding a beam FWHM of ~5 μm.

**Numerical Calculations.** The field enhancement of gap plasmons was assessed using commercial FDTD (finite-difference time-domain) software. Scanning electron microscopy images of the gold metasurface were imported into the program, where gold's permittivity was modeled as a dispersive Drude metal based on parameters fitted from the data provided by Palik. The simulation utilized a plane wave incident at normal incidence, with a mesh size set to 0.5 nm in in-plane (both *x*- and *y*-directions) and 0.2 nm out-of-plane (in the *z*-direction), ensuring that the results converged effectively.

**Theoretical Approach.** We employ a fully microscopic finite-temperature theory to describe the optical properties of FPs[19]. In this framework, the exciton self-energy $\Sigma(\omega)$ in Eq. (2) is expressed in terms of the electron-exciton *T*-matrix $\mathcal{T}(\mathbf{q},\omega)$ as

$$\Sigma(\omega) = \sum_{\mathbf{q}} f_{\boldsymbol{q}}\, \mathcal{T}\left(\mathbf{q}, \omega + \epsilon_{\mathbf{q}}\right) \tag{6a}$$

$$\mathcal{T}^{-1}(\mathbf{q},\omega) = -\frac{1}{v} - \sum_{\mathbf{k}} \frac{1-f_{\boldsymbol{k}}}{\omega-\epsilon_{\mathbf{k}}-\omega_{\mathbf{k}-\mathbf{q}}+i0}, \tag{6b}$$

where $f_{\boldsymbol{k}} = [e^{\frac{\epsilon_{\mathbf{k}}-\mu}{k_B T}} + 1]^{-1}$ is the electron Fermi-Dirac distribution at temperature $T$ and chemical potential $\mu = T \ln\,(e^{\frac{E_F}{k_B T}} - 1)$. The electron and exciton dispersions are $\epsilon_{\mathbf{k}} = k^2/2m$ and $\omega_{\mathbf{k}} = k^2/2m_X$, respectively. $v = (\sum_{\mathbf{k}} \frac{1}{\varepsilon_T+\epsilon_{\mathbf{k}}+\omega_{\mathbf{k}}})^{-1}$ is the electron-exciton interaction strength, where $\varepsilon_T$ is the trion binding energy.

The self-energy $\Sigma(\omega)$ fully determines the exciton Green's function (Eq. (2)) and, consequently, the optical response of the system, including absorption and PL as described by Eq. (1). Details of the numerical implementation are provided in Ref.[19], where particular care is required in treating spectral broadening. In the present work, experimentally extracted parameters are used, including $\varepsilon_T = 33$ meV and an exciton broadening $\eta_X = 30$ meV.

As described in the main text, comparing between calculated and measured temperature dependence of the energy splitting between the attractive and repulsive polaron branches enables full characterization of the FP properties, including the electron Fermi energy $E_F$, polaron energies $E_{A,R}$, spectral weights $Z_{A,R}$, and linewidths $\Gamma_{A,R}$. For monolayer $WS_2$ at room temperature, the thermal energy $k_B T = 26$ meV exceeds the Fermi energy $E_F = 5$ meV. In this regime, the AP is not a well-defined quasiparticle resonance[19,20], i.e., $E_A \neq \Sigma(E_A)$. Consequently, the absorption $A(\omega)$ and PL spectra strongly deviate from Lorentzian line shapes[11,19,20]. Nevertheless, the FP framework remains valid across this regime and accurately captures the relevant spectral features. The theory describes the crossover from a quantum-degenerate regime with well-defined quasiparticles ($E_F \gtrsim k_B T$) to an incoherent trion–hole continuum regime ($E_F \lesssim k_B T$), as established in Ref.[19] and confirmed in temperature-dependent PL experiments[20,59]. For the monolayer sample considered here, the relatively low doping places the system in the trion-hole contimuum regime, whereas multilayer samples (2L and 4L) are expected to approach the crossover region and possibly enter the quasiparticle regime. Note that, thermal broadening is incorporated within the theory, while additional quantum dephasing effects for the repulsive branch because of decay to the attractive branch are neglected[28].

In the strong-coupling regime, hybridization with plasmonic mode further enhances the visibility of spectral features. Accordingly, the simplified form of the exciton Green's function in Eq. (5) remains a reliable and practical approximation for extracting FPP branch energies from the measured scattering spectra, consistent with Ref[19].

**Fitting Procedures.** The fitting strategy for dark-field scattering spectra depends on whether the middle polariton (MP) branch is spectrally resolved. Accordingly, different procedures are used for monolayer (1L) and multilayer (2L and 4L) $WS_2$.

For 1L $WS_2$, the MP branch is not spectrally resolved. The intrinsic FP parameters ($E_{A,R}$, $Z_{A,R}$, $\Gamma_{A,R}$) are first determined independently from PL-based microscopic analysis, while the plasmon parameters ($E_{pl}$, $\eta_{pl}$) are obtained from the bare metasurface spectra. The remaining fitting primarily refines the residual optical weights $x$ and $y$, while the extracted value of the coupling strength $\Omega$ can be independently extracted from the anticrossing of the polariton modes. The fitting is performed by minimizing the deviation between theory and experiment for four observables extracted from the scattering profile $S(\omega)$: the energy splitting $E_{UP} - E_{LP}$, the scattering intensity difference $S(E_{UP}) - S(E_{LP})$, the scattering intensity at the repulsive-polaron energy $S(E_R)$, and the contrast $\max[S(E_{LP,UP})] - S(E_R)$, where $E_{LP}$ and $E_{UP}$ are identified as maxima of $S(\omega)$. The resulting fits are shown in Figure 2e,f.

Because conventional PL does not independently provide branch-resolved FP parameters in these samples, the multilayer spectra are analyzed using the same validated scattering framework as a constrained phenomenological model. The coupling strength $\Omega$ and detuning $\delta = E_{pl} - E_R$ can also be independently extracted by the experimentally observed three-branch anticrossing and the measured plasmon resonance (Figure S7). The fitting procedure minimizes deviations in five experimentally accessible quantities: the energy splitting $E_{UP} - E_{LP}$, the intensity difference $S(E_{UP}) - S(E_{LP})$, the minimum scattering intensity at the attractive- and repulsive-polaron energies $\min[S(E_{A,R})]$, the contrast $\max[S(E_{LP,UP})] - \min[S(E_{A,R})]$, and the relative dip depth $S(E_A) - S(E_R)$. The corresponding fits are presented in Figure 3a,b and Figure 4b.

ASSOCIATED CONTENT

**Supporting Information**.

The Supporting Information is available free of charge on the ACS Publications website.

Synthesis route and properties of the conducting redox polymer substrate; scanning electron microscopy images of the 20 nm-thick gold multi-singular metasurface; temperature dependence of PL profiles: comparison with theory; plasmonic resonance of the bare Au metasurface; comprehensive analysis of Fermi polaron properties in monolayer $WS_2$ via PL and scattering spectra; detuning dependence of FPPs peaks; simplified coupled oscillator model and alternative fitting procedure; strain-tunable repulsive-to-attractive polaron conversion; surface plasmon polaritons supported by smooth gold films (PDF)

AUTHOR INFORMATION

Corresponding Author

Francisco J. Garcia Vidal; Email: fj.garcia@uam.es

Mengxiao Chen; Email: mengxiaochen@zju.edu.cn

Qi Jie Wang; Email: qjwang@ntu.edu.sg

Yu Luo; Email: yu.luo@nuaa.edu.cn

Author Contributions

[#]T. Wu and F. M. Marchetti contribute equally to the article. T. Wu conceived the research project. T. Wu and M. Chen fabricated the CRP substrates with varying electron densities. L. Liu and T. Wu prepared the $WS_2$ flakes and T. Wu integrated them with the flexible plasmonic metasurfaces. T. Wu conducted the optical experiments, while Z. Wang and T. Wu captured the SEM images. T. Wu performed the finite-difference time-domain simulations. F. M. Marchetti, A. Tiene, and T. Wu carried out the theoretical analysis. F. J. Garcia-Vidal, M. Chen, Q. Wang,

and Y. Luo supervised the research. T. Wu, F. M. Marchetti, A. Tiene, F. J. Garcia-Vidal, Yu Luo and Antonio I. Fernández-Domínguez analyzed the data. T. Wu drafted the manuscript. All authors contributed to data interpretation and manuscript editing. All authors have given approval to the final version of the manuscript.

Funding Sources

Distinguished Professor Fund of Jiangsu Province (Grant No. 1004-YQR24010); Fundamental Research Funds for the Central Universities, NUAA (No. NE2024007); National Research Foundation Singapore programme (NRF-CRP29-2022-0003 and MSG-2023-0002); A*STAR grant numbers (M23M2b0056); Singapore Ministry of Education Academic Research Fund Tier 3 (MOET32024-0001); Spanish Ministry of Science, Innovation and Universities through the "Maria de Maetzu" Programme for Units of Excellence in R\&D (CEX2023-001316-M); Comunidad de Madrid and the Spanish State through the Recovery, Transformation and Resilience Plan ["MATERIALES DISRUPTIVOS BIDIMENSIONALES (2D)" (MAD2D-CM)-UAM7]; Proyecto Sinérgico CAM 2020 Y2020/TCS-6545 (NanoQuCo-CM); Ministerio de Ciencia e Innovación (MICINN), project No.~AEI/10.13039/501100011033 PID2020-113415RB-C22 (2DEnLight).

Notes

The authors declare no competing financial interest.

ACKNOWLEDGMENT

We acknowledge fruitful discussions with M.M. Parish and J. Levinsen. Y. Luo acknowledge financial support from Distinguished Professor Fund of Jiangsu Province (Grant No. 1004-YQR24010) and Fundamental Research Funds for the Central Universities, NUAA (No. NE2024007). Q. J. Wang acknowledges financial support from National Research Foundation

Singapore programme (NRF-CRP29-2022-0003 and MSG-2023-0002), A*STAR grant numbers (M23M2b0056), and Singapore Ministry of Education Academic Research Fund Tier 3 (MOET32024-0001). FMM, AIFD, and FJGV acknowledge financial support from the Spanish Ministry of Science, Innovation and Universities through the "Maria de Maetzu" Programme for Units of Excellence in R\&D (CEX2023-001316-M), from the Comunidad de Madrid and the Spanish State through the Recovery, Transformation and Resilience Plan ["MATERIALES DISRUPTIVOS BIDIMENSIONALES (2D)" (MAD2D-CM)-UAM7], and from the Proyecto Sinérgico CAM 2020 Y2020/TCS-6545 (NanoQuCo-CM). FMM acknowledges financial support from the Ministerio de Ciencia e Innovación (MICINN), project No.~AEI/10.13039/501100011033 PID2020-113415RB-C22 (2DEnLight).

Supplementary Materials for

# Room-temperature tuning and probing of Fermi polarons in atomically thin semiconductors on a plasmonic metasurface

Tingting Wu[1,#], Francesca Maria Marchetti[2,3,#], Antonio Tiene[2,3], Antonio I Fernández-Domínguez[2,3], Miao Qi[1], Zhe Wang[4], Lin Liu[1], Lei Wei[1], Francisco J. Garcia Vidal[2,3,*], Mengxiao Chen[5,*], Qi Jie Wang[1,*] and Yu Luo[6,*]

[1]School of Electrical and Electronic Engineering, Nanyang Technological University, Singapore 639798, Singapore

[2]Departamento de Física Teórica de la Materia Condensada, Universidad Autónoma de Madrid, Madrid 28049, Spain

[3]Condensed Matter Physics Center (IFIMAC), Universidad Autónoma de Madrid, Madrid 28049, Spain

[4]Key Laboratory of Bionic Engineering (Ministry of Education), Jilin University, Changchun 130022, China

[5]Department of Biomedical Engineering, Key Laboratory for Biomedical Engineering of Education Ministry, Zhejiang University, Hangzhou 310027, China

[6]National Key Laboratory of Microwave Photonics, Nanjing University of Aeronautics and Astronautics, Nanjing 211106, China

[#]These authors contributed equally: Tingting Wu, Francesca Maria Marchetti

[*]emails: fj.garcia@uam.es; mengxiaochen@zju.edu.cn; qjwang@ntu.edu.sg; yu.luo@nuaa.edu.cn

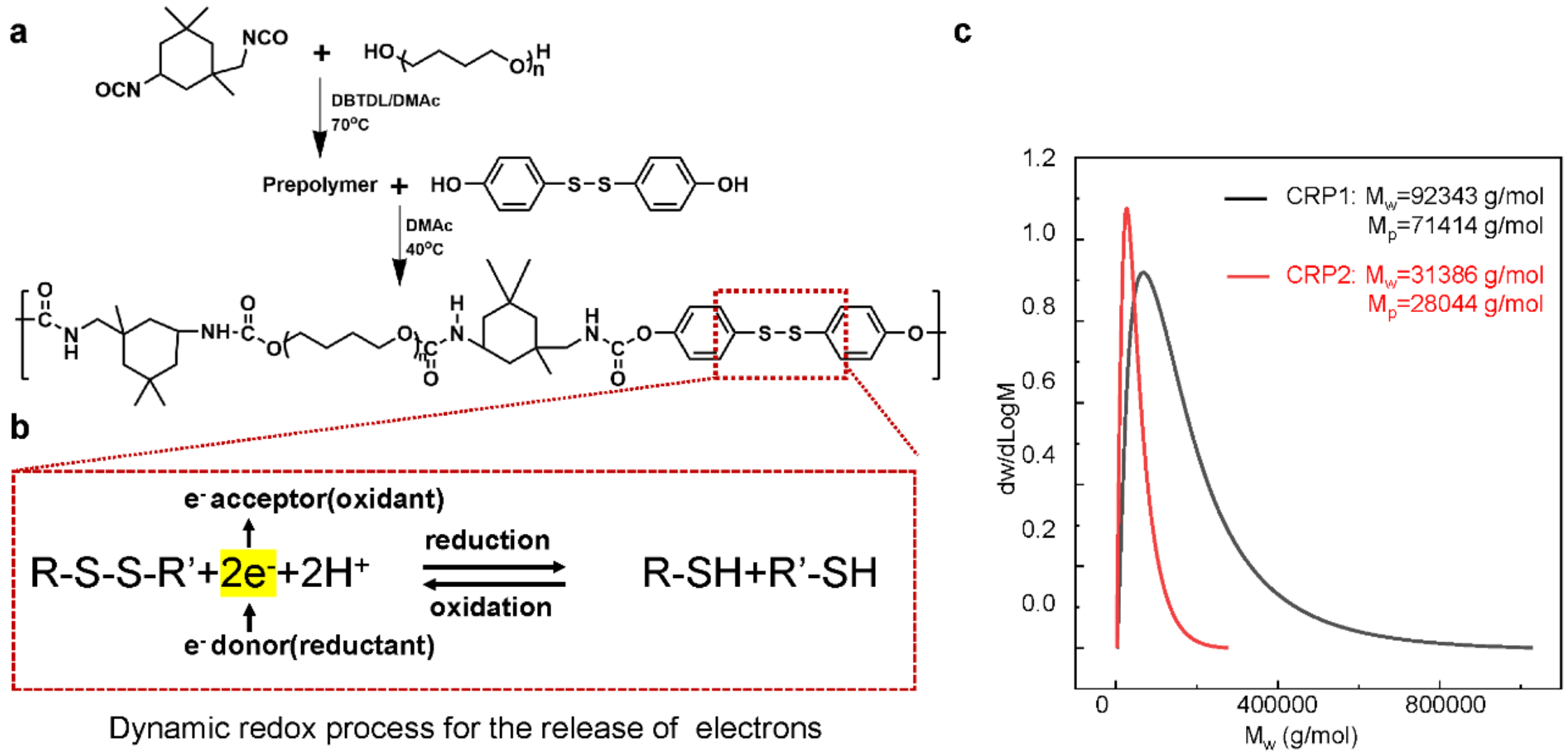

**Figure S1. Synthesis route and properties of the conducting redox polymer (CRP) substrate, which generates free electrons through redox reactions[1-4].** (**a**) Synthesis route of the CRP. (**b**) Dynamic redox process of the dynamic disulfide bonds. (**c**) Molecular weight distribution of the prepared CRP1 and CRP2 substrates. CRP1: weight average molecular weight $M_w$=92343 g/mol, peak average molecular weight $M_p$=71414 g/mol; CRP2: weight average molecular weight $M_w$=31386 g/mol, peak average molecular weight $M_p$=28044 g/mol. Lower molecular weight polymers (CPR2) exhibit greater chain mobility, allowing the disulfide bonds (covalent bonds formed between two sulphur atoms that can reversibly form and break in response to environmental conditions) in the CRP to undergo more frequent bond exchange and reformation[5,6]. This increased mobility enhances electron transfer and increases electronic density. In contrast, higher molecular weight polymers (CPR1) exhibit reduced chain mobility, which limits the dynamic behavior of the disulfide bonds, resulting in slower electron transfer and reduced electronic density[7].

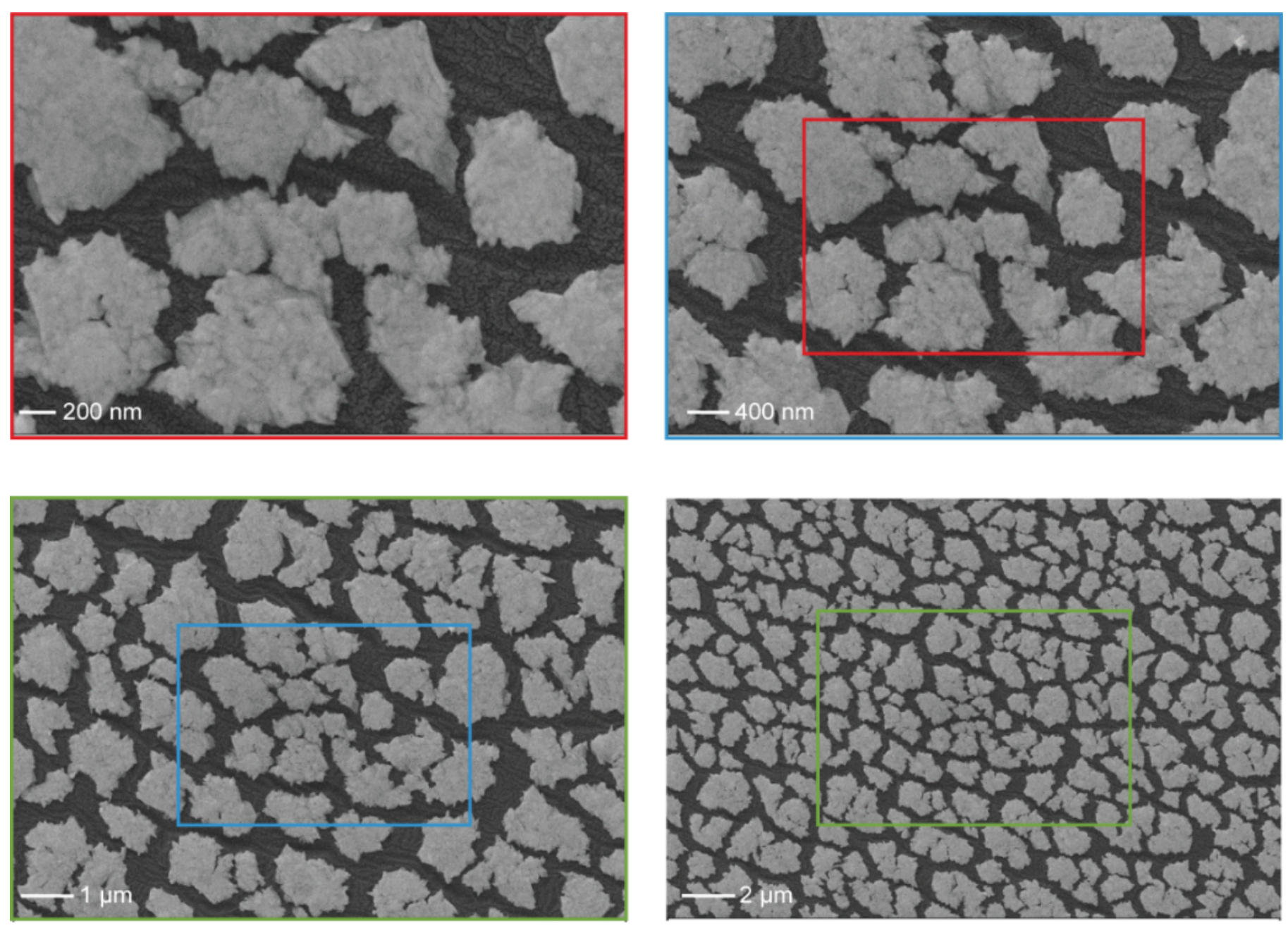

**Figure S2. Scanning electron microscopy images of the 20 nm-thick gold multi-singular metasurface, revealing densely packed nanoscale plasmonic gaps**. From the first to the fourth image, each boxed region indicates the area that is shown at higher magnification in the subsequent image.

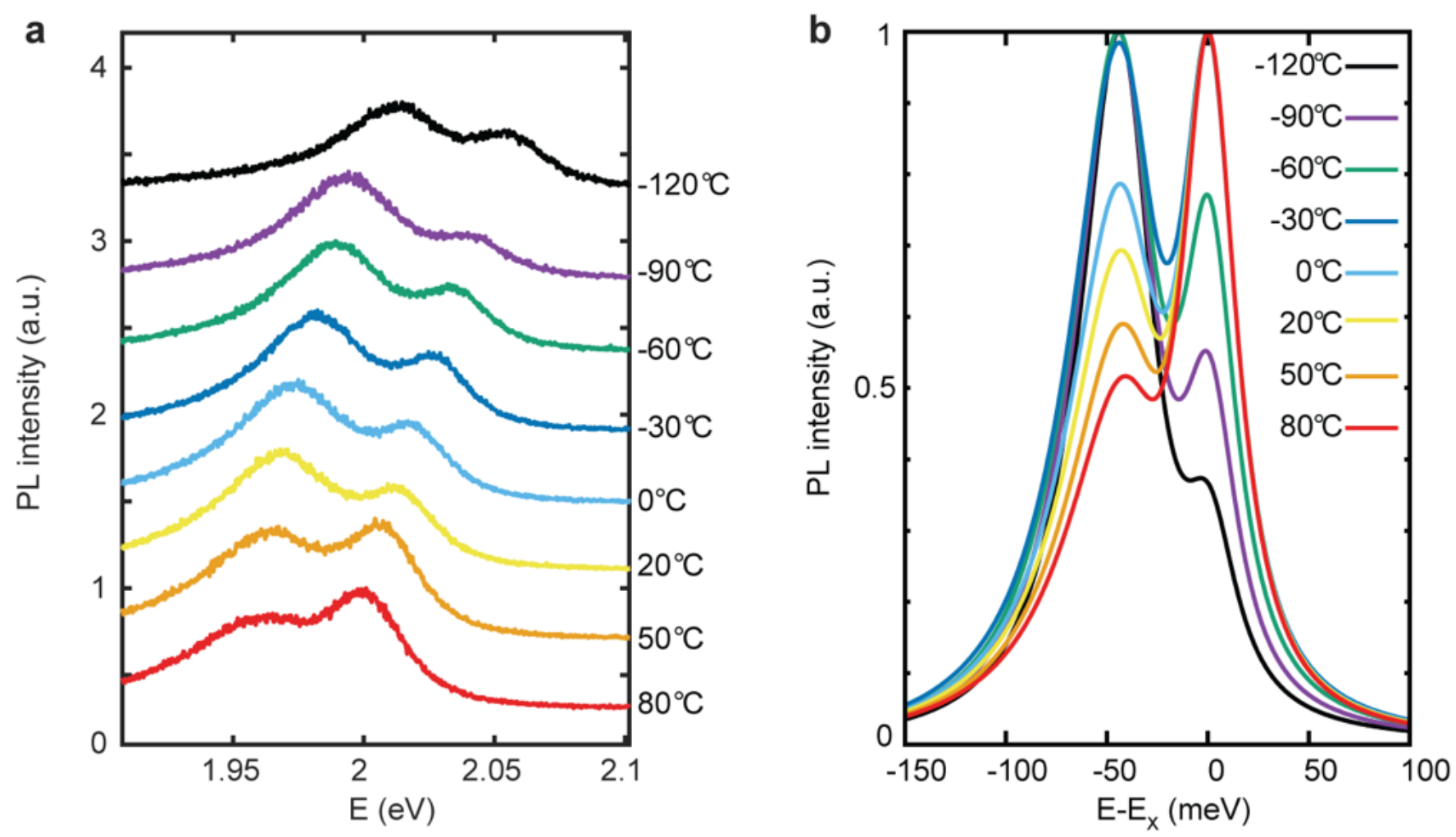

**Figure S3. Temperature dependence of PL profiles: comparison with theory.** (**a**) PL emission spectra of monolayer $WS_2$ on the CRP1 substrate at different temperatures. (**b**) Theoretical PL profiles for Fermi polarons at different temperatures at a constant electron density of $E_F \sim 5$ meV. The temperature dependence of the peak splitting shows a quantitative agreement between theory and experiments (see Figure 2c in the main text). Furthermore, the relative PL intensities of the two peaks are in qualitative agreement, showing a shift in intensity from the attractive branch to the repulsive branch with increasing temperature due to changes in occupation --- see Eq. (1) of the main text.

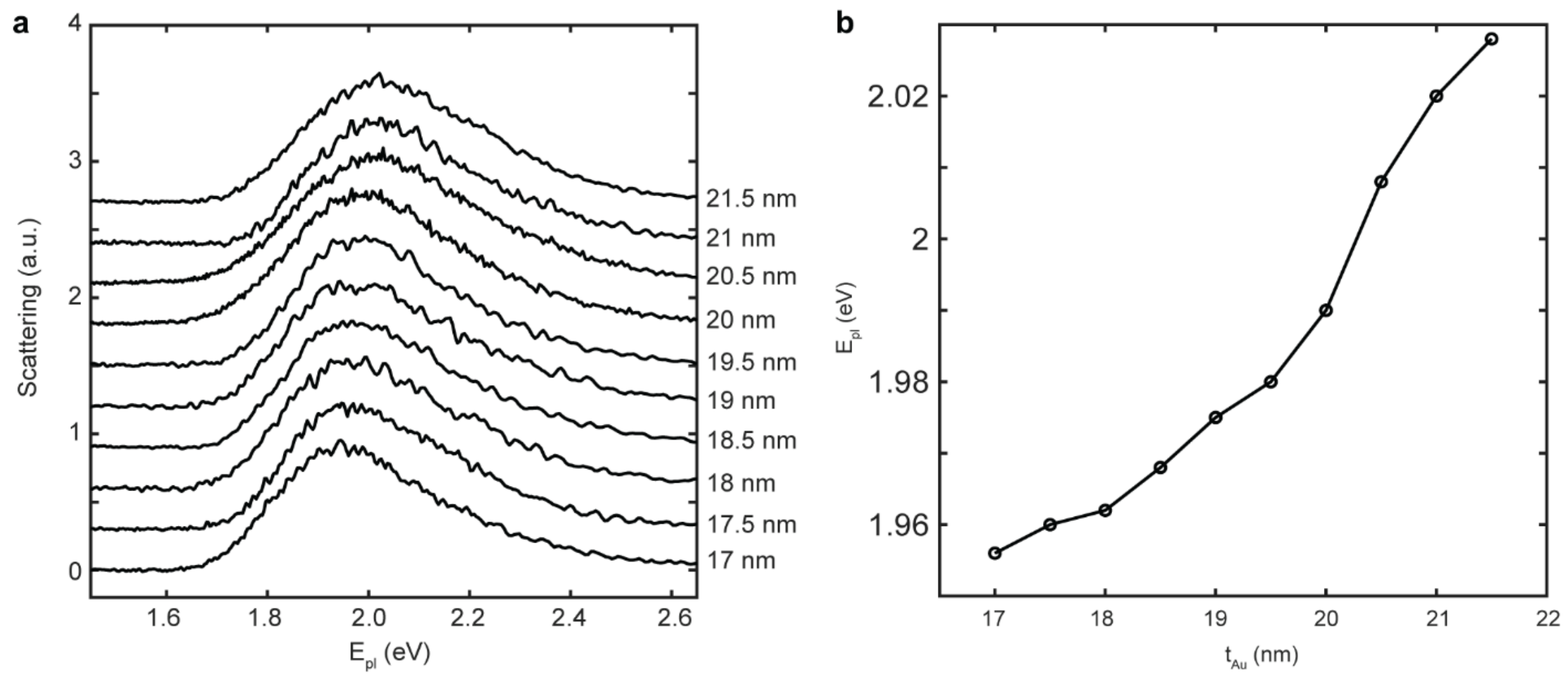

**Figure S4. Plasmonic resonance of the bare Au metasurface as a function of Au thickness. (a)** Dark-field scattering spectra (arbitrary units, a.u.) of the Au metasurface measured in the absence of $WS_2$ for Au thicknesses ranging from 17.0 nm to 21.5 nm. (**b**) Extracted plasmonic resonance energy as a function of Au thickness.

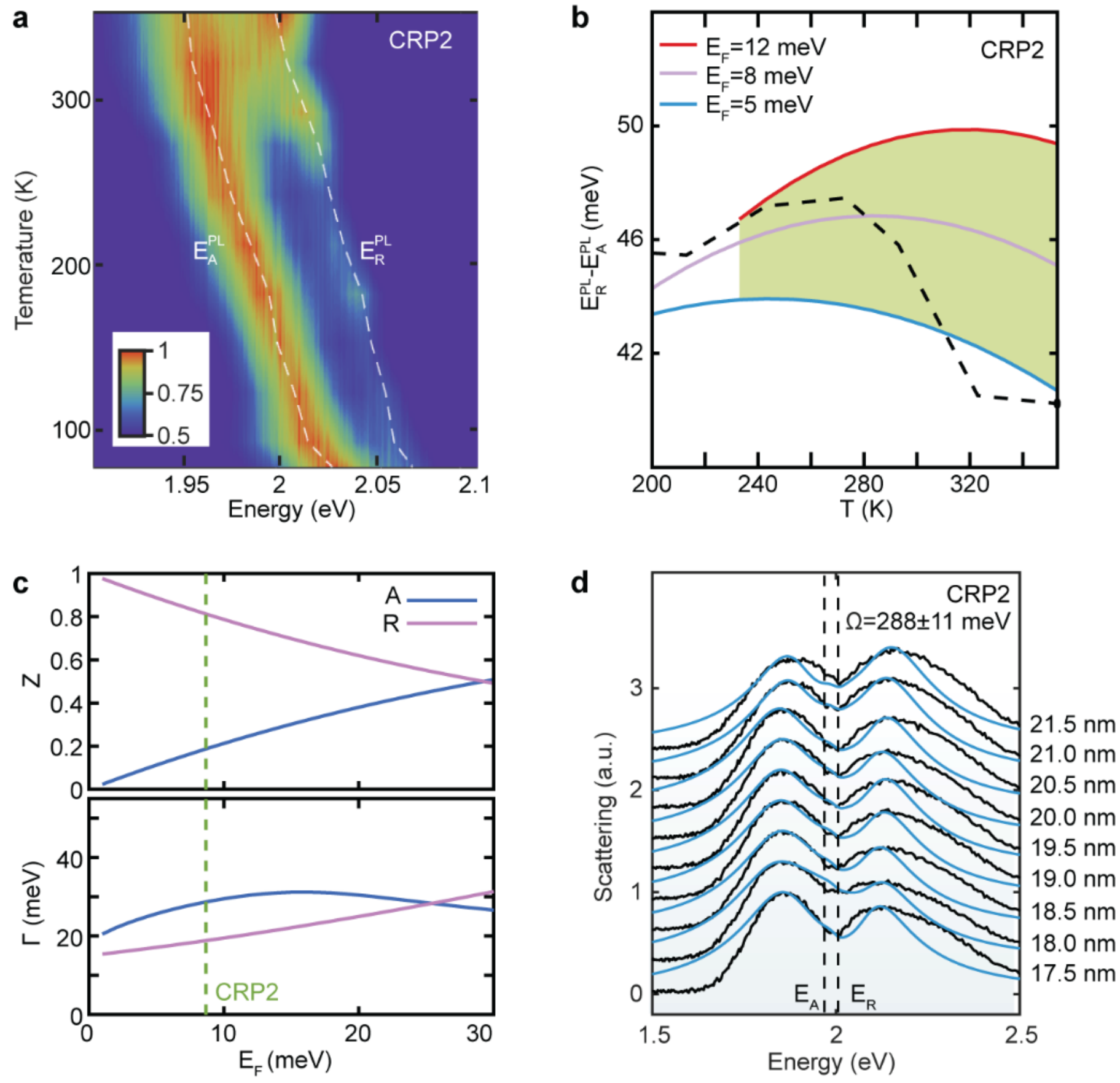

**Figure S5. Comprehensive analysis of Fermi polaron properties in monolayer $WS_2$ via PL and scattering spectra using CRP2 polymer substrate.** (**a**) PL emission mapping of monolayer $WS_2$ on CRP2 substrate at different temperatures. (**b**) Temperature-dependent energy splitting between attractive (A) and repulsive (R) polaron peaks, $E_R^{PL} - E_A^{PL}$, measured in PL (dashed line), showing consistent behavior with theoretical modeling (solid lines) in a limited interval of doping levels ($E_F$). (**c**) Expected doping dependence of spectral weights ($Z$) and half-linewidths ($\Gamma$) and specific values (vertical dashed lines) extracted from the PL spectra. (**d**) Dark-field scattering spectra of monolayer $WS_2$ on Au metasurface with CRP2 substrate over different Au film thicknesses (increasing from 17.5 nm to 21.5 nm from bottom to top). Vertical dashed lines mark the $WS_2$ FP energies, $E_{A,R}$ extracted from PL at weak coupling ($E_{A,R}^{PL}$ and $E_{A,R}$ differs because of temperature effects). The black and blue lines show the measured and calculated scattering spectra,

respectively, determined using the $Z_{A,R}$ and $\Gamma_{A,R}$ values from the weak coupling PL spectra and fitting the remaining parameters.

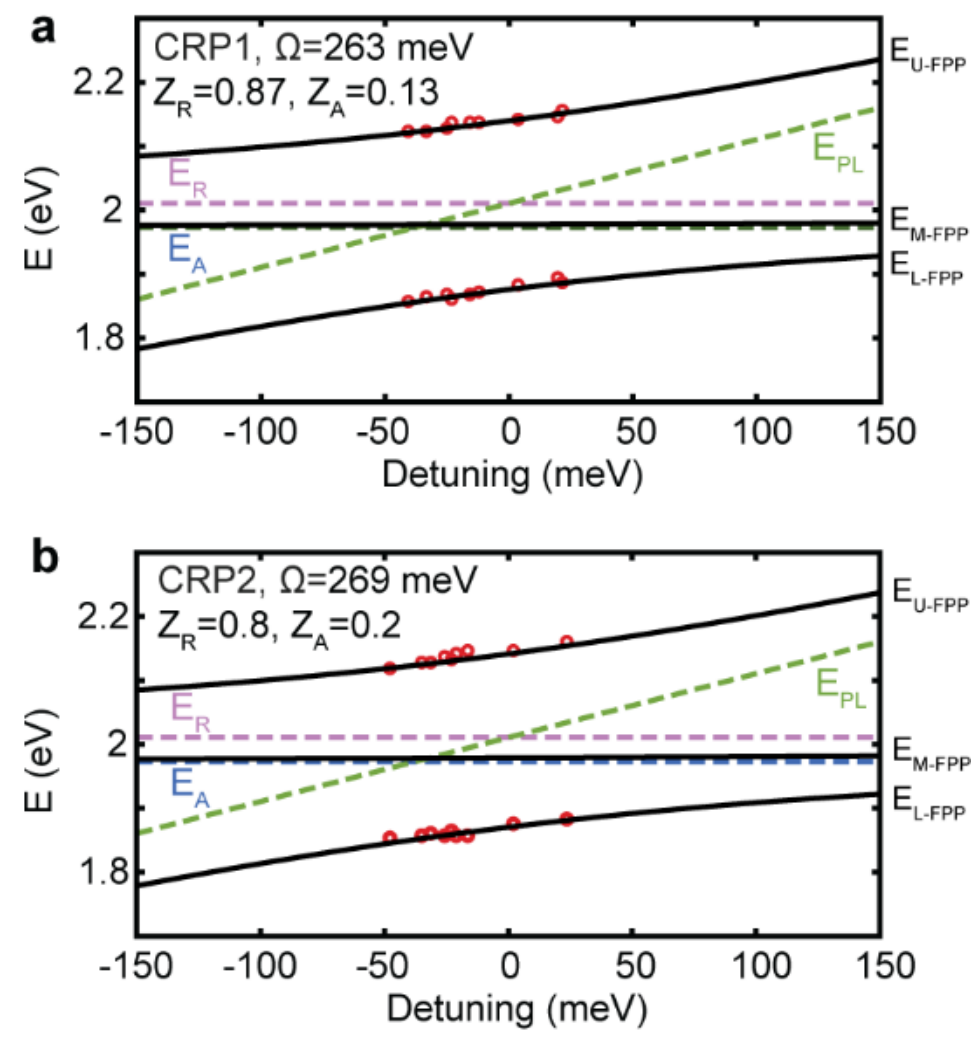

**Figure S6. Detuning dependence of L-FPP and U-FPP peaks for monolayer $WS_2$ and fitting with eigenvalues of the coupled oscillator model.** FPP refers to Fermi polaron polariton. (**a**), (**b**), Detuning dependence of FPPs energies obtained in the dark-field scattering spectra of Figure 2e (**a**) and Figure 2f (**b**) in the main text. The lower ($E_{L-FPP}$) and upper ($E_{U-FPP}$) polariton energies extracted from the spectra (red symbols) are fitted (solid black lines) with the eigenvalues of a coupled oscillator model (Methods) giving $\Omega = 263$ meV (CRP1 in (**a**) and 269 meV (CRP2 in (**b**)). Dashed lines are the uncoupled polaron energies ($E_{A,R}$) and the plasmonic mode energy ($E_{pl}$). For monolayers, no middle polariton (M-FPP) peak could be measured because the spectral weight of the repulsive polaron is much larger than the attractive polaron. This implies that the M-FPP energy $E_{M-FPP}$ is very close to the attractive branch energy $E_A$, making it difficult to resolve due to broadening effects.

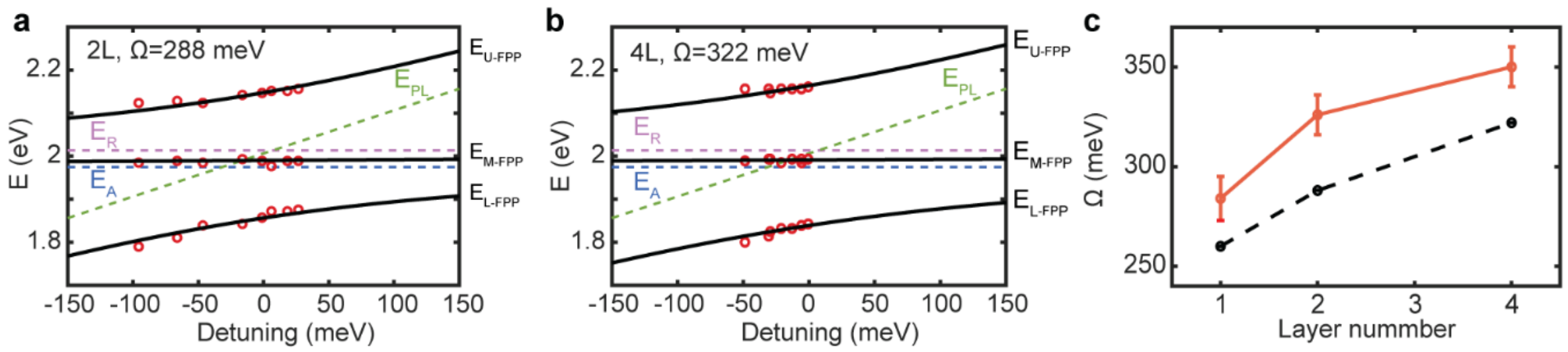

**Figure S7. Detuning dependence of FPPs peaks for multilayer $WS_2$ and fitting with eigenvalues of a coupled oscillator model.** Detuning dependence of FPPs energies obtained in the dark-field scattering spectra for (**a**) bilayer (2L) and (**b**) quadrilayer (4L) $WS_2$. The lower ($E_{L-FPP}$), middle ($E_{M-FPP}$) and upper ($E_{U-FPP}$) polariton energies extracted from the spectra (red symbols) are fitted (solid black lines) with the eigenvalues of a coupled oscillator model (Methods) giving $\Omega = 288$ meV (2L) and 322 meV (4L). Dashed lines are the uncoupled polaron energies ($E_{A,R}$) and the plasmonic mode energy ($E_{pl}$). (**c**) The Rabi splitting ($\Omega$) as a function of the $WS_2$ layer number, derived from the eigenvalues of a three-coupled oscillator model (black dots and dashed line), showing compatible values of the Rabi coupling extracted from our developed fully microscopic model (red dots and solid lines, see Figure 3c in the main text for details).

**Simplified coupled oscillator model and alternative fitting procedure.**

In strong coupling conditions with metasurface plasmons, we can extract the dark scattering spectrum (Eq. (3) in the main text) describing FPPs, which can be compared to experiments by employing a simplified description of the exciton Green's function (Eq. (5) in the main text). This is equivalent to consider the following three coupled oscillator model (COM) Hamiltonian

$$H = \begin{pmatrix} E_A - i\Gamma_A & 0 & \Omega_A/2 \\ 0 & E_R - i\Gamma_R & \Omega_R/2 \\ \Omega_A/2 & \Omega_R/2 & E_{pl} - i\eta_{pl} \end{pmatrix}, \tag{S1}$$

where $\Omega_{A,R} = \Omega\sqrt{Z_{A,R}}$ and thus the overall Rabi coupling is $\Omega = \sqrt{\Omega_A^2 + \Omega_R^2}$. One can easily show then that $G_{pl}(\omega) = (\omega - H)^{-1}_{33}$. The metasurface plasmon broadening is $\eta_{pl} = 380$ meV.

As an alternative, albeit less accurate, fitting procedure, we compare the fitted values of the overall Rabi splitting $\Omega$ obtained by fitting the entire scattering profile (see main text) with those obtained using the eigenvalues of the three-COM (Eq. S1). For 2L and 4L, we determine the detuning $\delta$ by considering the trace of $H$ (neglecting the contribution of the imaginary parts):

$$E_{LP} + E_{MP} + E_{UP} = E_A + E_R + E_{pl} = E_A + 2E_R + \delta, \tag{S2}$$

while for 1L, we neglect the contribution of the MP peak which cannot be resolved ($E_{LP} + E_{UP} = E_R + E_{pl} = 2E_R + \delta$).

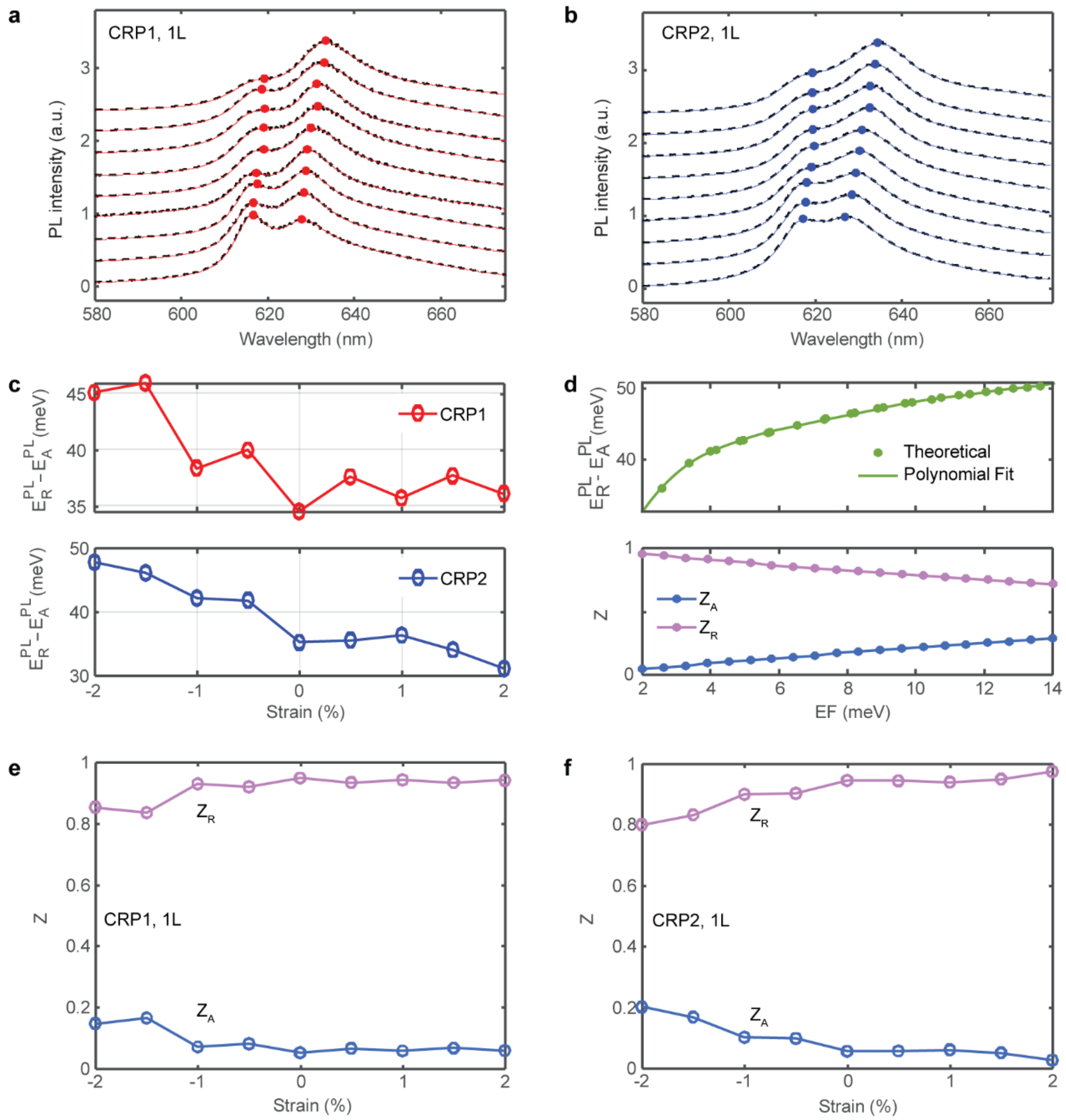

**Figure S8. Strain-tunable repulsive-to-attractive polaron conversion in plasmon-integrated monolayer $WS_2$ systems.** PL spectra of monolayer $WS_2$ on Au metasurfaces under varying uniaxial strain, ranging from +2% to -2% (top to bottom), on CRP1 (**a**) and CRP2 (**b**) polymer substrates. **c**, Energy splitting between repulsive (R) and attractive (A) polaron peaks, $E_R^{PL} - E_A^{PL}$, extracted from PL measurements (upper panel: CRP1 with red dots marking PL polaron peaks in (**a**); lower panel: CRP2 with blue dots marking PL polaron peaks in (**b**), demonstrating strain-

dependent tunability. (**d**) Upper: theoretically calculated values of $E_R^{PL} - E_A^{PL}$ (dots) with polynomial fit (lines). Lower: theoretically calculated spectral weight (Z) as a function of doping. (**e**), (**f**), Extracted spectral weights as a function of applied uniaxial strain.

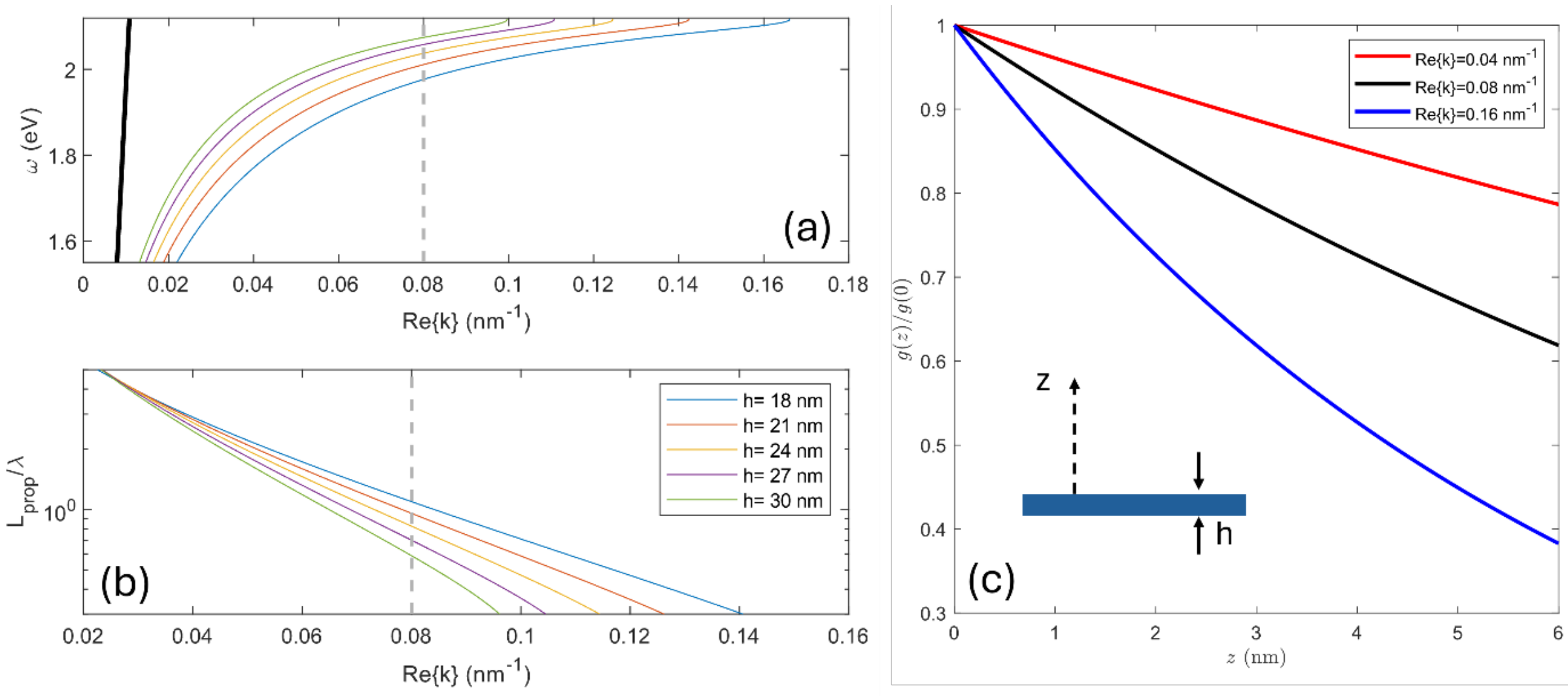

**Figure S9. Surface plasmon polaritons supported by smooth gold films of varying thickness**. Surface plasmon polaritons sustained by smooth Au films of different thickness, h, described in the quasi-static approximation[8], $k(\omega) = \frac{1}{h}\mathrm{Ln}\left(\frac{\epsilon_{\mathrm{Au}}(\omega)-\epsilon_{\mathrm{bg}}}{\epsilon_{\mathrm{Au}}(\omega)+\epsilon_{\mathrm{bg}}}\right)$. The gold permittivity, $\epsilon_{\mathrm{Au}}(\omega)$ is a Drude fitting to gold permittivity[9] and $\epsilon_{\mathrm{bg}} = 3$ accounts for the TMD and substrate in the film background. (a) Dispersion relation, frequency, $\omega$, versus real part of the propagating wave-vector, $\mathrm{Re}[k]$. (b) Propagation length, $L_{\mathrm{prop}} = (2\,\mathrm{Im}[k])^{-1}$ versus $\mathrm{Re}[k]$. Vertical dashed lines mark the condition $\mathrm{Re}[k] = 0.08\ \mathrm{nm}^{-1}$, for which $L_{\mathrm{prop}} \approx \lambda \approx 600$ nm (typical size of the islands in the plasmonic measurface in the experimental setup). (c) Exponential decay (within a few nm) of the light-matter strong coupling constant associated to the surface plasmon polaritons, $g(z) = g(z)e^{-\mathrm{Re}[k]z}$ for wave-vector variations around $\mathrm{Re}[k] = 0.08\ \mathrm{nm}^{-1}$.

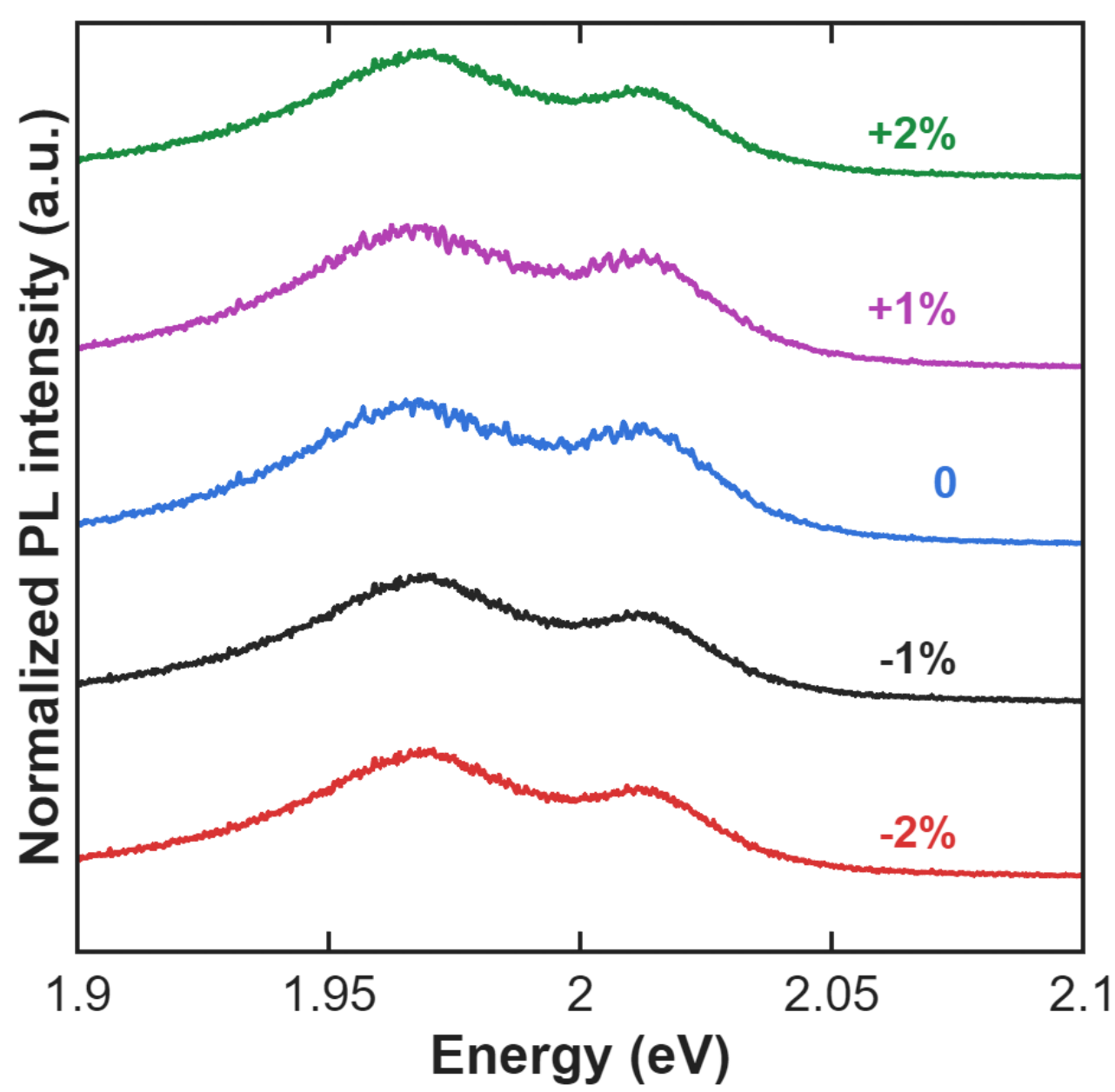

**Figure S10. Strain-dependent PL emission spectra of electron-doped monolayer $WS_2$ on Au metasurfaces, with a 20 nm $Al_2O_3$ spacer inserted between $WS_2$ and the Au metasurface**. The spacer suppresses plasmon–polaron coupling and hybridization, resulting in negligible peak-energy shifts, minor linewidth variations, and no discernible spectral-weight transfer between branches.